\documentclass[aps,superscriptaddress,showpacs,twocolumn]{revtex4-1}

\usepackage{graphicx}
\usepackage{color}
\usepackage{graphicx}
\usepackage{amssymb}
\usepackage{epstopdf}
\usepackage{times}
\usepackage{ulem}
\usepackage{footnote}
\begin{document}


\title{Highly p-doped graphene obtained by fluorine intercalation} 
\date{\today}


\author{Andrew L. Walter$^{1,2,\footnotemark}$, Ki-Joon Jeon$^{3}$, Aaron Bostwick$^{1}$, Florian Speck$^{4}$, Markus Ostler$^{4}$, Thomas Seyller$^{4}$, Luca Moreschini$^{1}$, Yong Su Kim$^{1,5}$, Young Jun Chang$^{1,2}$, Karsten Horn$^{2}$, Eli Rotenberg}

\address{ Advanced Light Source (ALS), E. O. Lawrence Berkeley National Laboratory, Berkeley, California 94720, USA.\\ $^{2}$ Department of Molecular Physics, Fritz-Haber-Institut der Max-Planck-Gesellschaft, Faradayweg 4-6, 14195 Berlin, Germany.\\ $^{3}$School of Electrical Engineering, University of Ulsan, Namgu, Ulsan, 680-749, South Korea\\ $^{4}$ Lehrstuhl f\"{u}r Technische Physik, Universit\"{a}t Erlangen-N\"{u}rnberg, Erwin- Rommel-Strasse 1, 91058 Erlangen, Germany.\\${5}$ Department of Applied Physics, Hanyang University, Ansan, Gyeonggi-do
426-791, Korea.}

\footnotetext{corresponding author: alwalter@lbl.gov}



\date{\today}

\def\EF{$E_{\mathrm{F}}$}
\def\ED{$E_{\mathrm{D}}$}
\def\kk{K-K}
\def\gk{$\Gamma$-K}
\def\gkm{$\Gamma$-K-M}
\definecolor{eli}{rgb}{0,0,0}
\def\eli{\textcolor{eli}}

\begin{abstract}
We present a \eli{method for decoupling epitaxial graphene grown on SiC(0001) by intercalation of a layer of fluorine at the interface}. The fluorine atoms do not enter into a \eli{covalent} bond with graphene, but rather saturate the substrate Si bonds. This configuration of the fluorine atoms induces a \eli{remarkably large} hole density of  $p  \approx 4.5 \times 10^{13}$ cm$^{-2}$, equivalent to the location of the Fermi level at 0.79 eV above the Dirac point $E_D$.

\end{abstract}


\maketitle 



Many applications of graphene, the single layer of hexagonally coordinated carbon atoms have been proposed  since the discovery of its unusual electronic properties \cite{Geim:2007p4020}. For these applications to be realized, \eli{the ability to achieve} $n$- and $p$-type doping is \eli{required}. While $n$-type doping is observed in graphene grown epitaxially on SiC \eli{\cite{Bostwick:2007p247,Emtsev:2009p212}},  only marginal $p$-type doping has been achieved by hydrogen\cite{Riedl:2009p4922,Speck:2010p5373,Bostwick:2010p3387,Virojanadara:2010p5374} and Au\cite{Gierz:2010p4973} intercalation or the adsorption of fluorine-containing molecules such as F4-TCNQ \cite{Coletti:2010p5365}. Here we present a technique for producing highly $p$-doped graphene on SiC via fluorine intercalation. In contrast to the previous work on fluorinated graphene, (theoretical \cite{Boukhvalov:2009p4022} and experimental \cite{Withers:2010p4016,Robinson:2010p4582,Nair:2010p5371,Jeon:2010p5372}) where the fluorine is bound to the graphene transforming it into an insulator, here the graphene retains the linear Dirac dispersion that leads to many of its unusual properties.
 
Previously, hydrogen \cite{Riedl:2009p4922,Speck:2010p5373,Bostwick:2010p3387,Virojanadara:2010p5374} and gold \cite{Gierz:2010p4973} have been shown to intercalate between layers of graphene and SiC(0001) produced by thermal decomposition. In both cases a slight $p$-doping of the graphene is observed, $p \approx 6\times 10^{12}$ cm$^{-2}$ and  $p  \approx 7\times10^{11}$ cm$^{-2}$ respectively.  \eli{Surface treatment with strongly oxidizing molecules such as NO$_2$ has yielded a large net $p$-doping change around $2\times 10^{13}$ cm $^{-2}$\cite{Zhou:2008p5400}, but these molecules are not thermally stable, and it is not clear how to protect the molecules from exposure to ambient conditions.}  In the present study we have used core level X-ray Photoemission Spectroscopy (XPS) and Angle Resolved Photoelectron Spectroscopy (ARPES) to investigate the interaction of fluorine atoms with the  ($6\sqrt{3} \times 6\sqrt{3}$) R30$^\circ$/SiC structure and to quantify the amount of $p$- doping. \eli{Properly treated, graphene becomes doped to an extremely high level ($p  \approx 4.5\times 10^{13}$ cm$^{-2}$), and is stable in air and at high temperatures. }

\begin{figure}
\includegraphics{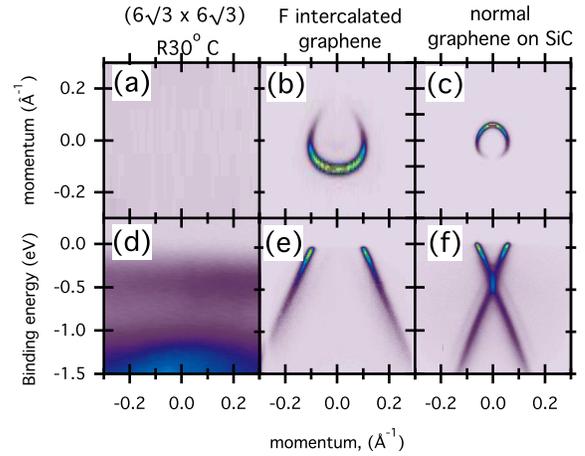}
\caption{\label{ARPES} Experimental constant energy cut at the Fermi energy (a,b and c) and valence band map  (d,e and f) from (a and d) \eli{buffer layer}, (b and e) graphene prepared by interaction of fluorine with the  \eli{buffer layer}, and (c and f) graphene on  \eli{buffer layer}. The $p$-doped character of the fluorine intercalated sample is evident. }
\end{figure}

\begin{figure}
\includegraphics{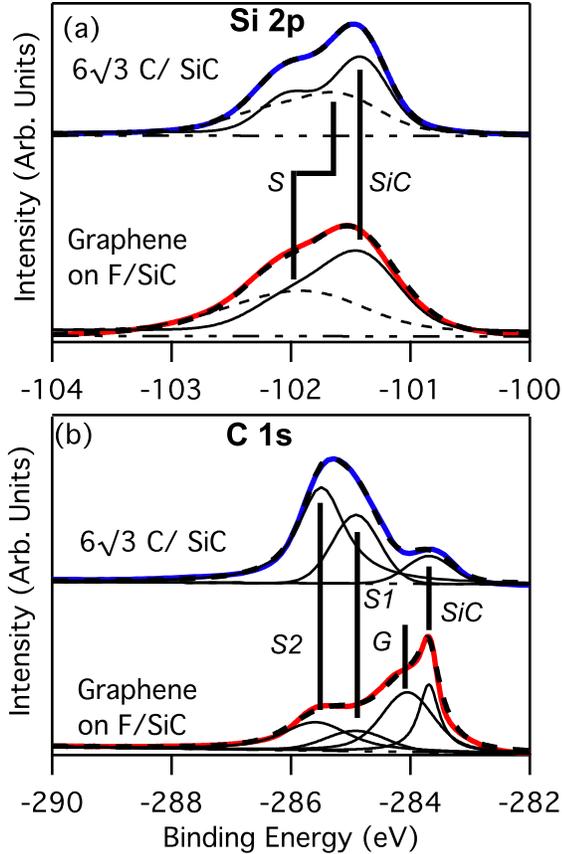}
\caption{\label{XPS} \eli{X-ray Photoemission Spectra (XPS) data taken using 350 eV photons of the ($6\sqrt{3} \times 6\sqrt{3}$) R30$^\circ$ buffer layer before (upper panels) and after (lower panels) fluorination. In each case the solid red and blue lines represent the data while the dashed black line represent the fit. (a) Si 2$p$ core levels are fitted to two spin-orbit-split doublet peaks, representing the bulk (SiC) and surface ($S$) contributions\cite{Carlisle:1994p5380}. Fluorine bonding to interfacial Si is indicated by a a shift in the surface doublet, $S$, of -0.18 eV  after flourination. (b) C 1$s$ core levels before fluorination consist of a bulk (SiC) peak and two surface peaks ($S_1, S_2$) representing buffer layer C atoms bonded to the substrate \cite{Emtsev:2008p4975}.   After fluorination, the latter peaks are weakened, and a new graphene-derived peak ($G$) is observed.  No C\textendash F bonds are observed.}}
\end{figure}

Experiments were performed on ($6\sqrt{3} \times 6\sqrt{3}$) R30$^\circ$ carbon ``buffer layer'' on a silicon terminated SiC(0001) surface, prepared by the method of Ostler et al. \cite{Ostler:2010p5375}. The carbon buffer layer is a covalently bound, \eli{electrically inactive} graphene layer\cite{Emtsev:2008p4975} in which about one third of the C atoms bind to the Si atoms on the SiC(0001) substrate. This bonding breaks up the $\pi$ band network so that the ($6\sqrt{3} \times 6\sqrt{3}$) R30$^\circ$ C layer \eli{(henceforth \textit{buffer layer})} has no \eli{D}irac cone.  Hydrogen and gold have been shown to intercalate underneath the \eli{buffer layer}, breaking the carbon bonds to the substrate and thus decoupling it from the SiC and transforming it into graphene \cite{Riedl:2009p4922,Speck:2010p5373,Bostwick:2010p3387,Virojanadara:2010p5374}. \eli{Samples were fluorinated} following a similar method to McFeely et al.\cite{Mcfeely:1984p5366} whereby  \eli{buffer layer} samples are placed in a reaction chamber with a XeF$_2$ crystal and heated to 200 $^{\rm o}$ C for 2 hours, leading to the dissociation of the XeF$_{2}$ and the liberation of reactive fluorine atoms. This method has previously been used to study the interaction between fluorine and silicon surfaces\cite{Mcfeely:1984p5366} as well as fluorine and graphene \cite{Withers:2010p4016,Robinson:2010p4582,Nair:2010p5371,Jeon:2010p5372} and \eli{normally leads to the formation of C\textendash F bonds with the graphene layer}. If a sacrificial piece of molybdenum is included in the reaction chamber, an entirely different reaction is promoted, whereby the F atoms bond not to the buffer layer C atoms, but instead to the interfacial Si atoms, as discussed below. At high pressure, XeF$_2$ reacts with the molybdenum to produce MoF$_6$ through the well-known reactions (depending on whether the Mo is clean or oxidized) \cite{Veyan:2010p5370,BURNS:1977p5369}:
\begin{eqnarray}
\mathrm{3XeF _2 + Mo} & \rightarrow & \mathrm{ MoF_6 (g)+3Xe (g) }\\
\mathrm{6XeF_2 +  2MoO_3} & \rightarrow & \mathrm{2MoF_6 (g) + 6 Xe (g) + 3O_2 (g)}
\end{eqnarray}
Following the fluorination process, the samples are covered by a layer of Mo (oxidized by air exposure) which is readily removed by sonication in ethanol.  The presence of this Mo residue indicates that it is the gaseous MoF$_6$, not XeF$_2$, that interacts with the graphene. Samples were transported through air to the Electronic Structure Factory endstation at beamline 7 of the Advanced Light Source, Lawrence Berkeley National Laboratory and annealed to $\sim$ 600 $^{\rm o}$C in vacuum to clean the surface for ARPES and core level measurements.  

The Fermi surfaces and bandstructures of three samples are presented in Fig. \ref{ARPES} :  (a and d) a buffer-layer sample ($6\sqrt{3} \times 6\sqrt{3}$) R30$^\circ$/SiC) without fluorination; (b and e) the same buffer-layer sample later treated with fluorine, and for comparison, (c and f) a graphene sample grown conventionally, consisting of an active graphene layer on top of the inactive buffer layer \cite{Emtsev:2008p4975}.  

\eli{As reported previously, the buffer layer sample is insulating \cite{Emtsev:2008p4975}, having no Fermi surface and having only a flat band appearing around 0.4 eV binding energy (see Fig.\ 1Ai, Bi).  After fluorination, the circular Fermi surface and linear bands of graphene appear.  Projection of the linear bands indicate a Dirac crossing well above the Fermi level, as expected for $p$-type doping, and extrapolating the band to the region above the Fermi edge locates it 0.79 eV above \EF. The Fermi surface size can be used \cite{Ohta:2006p260} to quantitatively evaluate the hole density at ( $p  \approx 4.5\times 10^{13}$ cm$^{-2}$) which is an order of magnitude larger than previously observed for H intercalated graphene ( $p  \approx 6\times 10^{12}$ cm$^{-2}$).  For comparison, the conventional Argon grown graphene on buffer layer on SiC samples\cite{Ostler:2010p5375} are $n$-type, with a much smaller charge density ($n\approx 1\times 10^{13}$ cm$^{-2}$) and Fermi surface (Fig.\ 1Aiii, Biii).}\eli{These results suggest that the buffer layer is transformed by fluorination into a $p$-type graphene. The linewidth of the ARPES $\pi$ bands at the Fermi level ($\sim$0.014 \AA$^{-1}$ ), which directly gives the electron mean free path from defect scattering, is comparable to that from the conventional graphene sample($\sim$0.009 \AA$^{-1}$ ) indicating a similar clean, defect-free graphene layer on the fluorine samples. We now discuss core-level photoemission measurements of samples (i) and (ii) which demonstrate that the fluorine intercalates under the buffer-layer / SiC interface.  This is completely different from fluorographene/fluorographane materials recently discovered, which are wide-bandgap semiconductors owing to covalent C\textendash F bond formation\cite{Nair:2010p5371,Peelaers:2011p5383,Jeon:2010p5372} .}

Core level photoemission spectra of the \eli{buffer} layer before and after exposure to fluorine are presented in Fig.  \ref{XPS}. After fluorine treatment, a clear F 1$s$ peak is observed  at 686.9 eV binding energy (not shown), confirming the presence of fluorine. Except for a small residual oxygen peak only peaks relating to C, F and Si are present in XPS scans up to a binding energy of 850 eV.The Si 2$p$ peak features (Fig. \ref{XPS} a) are comparable to previous data from the  \eli{buffer} layer\cite{Seyller:2006p4029} and are fitted to two doublets representing the bulk (SiC) and surface, ($S$) contributions to the spectrum. The doublets consist of 2 Lorentzian-Gaussian peaks with identical line widths separated by -0.605 eV and with an amplitude ratio of 2:1. In all cases the Lorentzian width is ~ 0.08 eV, while both the bulk (0.5 to 0.7) and surface (0.75 to 1) gaussian widths increase after flourination. The increase in the Gaussian width is an indication of  \eli{inhomogeneity of the sample interface.}   The position of the bulk doublet is unchanged (-101.42) upon exposure to fluorine\eli{,  while} the surface doublet, $S$, shifts from -101.6 eV to -101.78 eV, showing that the fluorine \eli{interacts with, or is near to,} the substrate silicon atoms. This splitting is similar to, but larger than, the Si 2p splitting observed in H-intercalated graphene\cite{Riedl:2009p4922}. The observed chemical shift is much smaller than observed for SiF species\cite{Mcfeely:1984p5366}, indicating a relatively weak charge transfer from the Si atoms.

That the fluorine atoms do not covalently bond with the graphene, \eli{but rather intercalate to the C-SiC interface} is also evident from C 1$s$ core level spectra shown in Fig. \ref{XPS}(b). The top spectrum is from the  \eli{buffer} layer before fluorination, and exhibits two peaks ($S_1,S_2$) which arise from the bonding of the carbon atoms in the  \eli{buffer} layer to the SiC substrate, at $\sim$ -285.55 eV and $\sim$ -284.75 eV \cite{Emtsev:2008p4975}, and a peak at $\sim$ -283.7 eV due to the bulk SiC. Upon exposure to fluorine, \eli{buffer layer peaks $S_1, S_2$} are largely suppressed, and a new peak ($G$) at $\sim$ -284.05 eV \eli{appears, attributed} to the formation of the graphene layer. Its separation to the carbon peak from the SiC bulk is much smaller than for graphene on  \eli{buffer}/SiC samples ($\sim$ -284.74 eV)\cite{Emtsev:2008p4975}, which is  related to the fact that the graphene is strongly doped. The C 1$s$ spectrum in Fig. \ref{XPS}(b, lower) also shows none of the C-F  peaks  \eli{of} perflourographene\cite{Robinson:2010p4582}, or \eli{any peaks in the range 285 eV to 290 eV expected for  C\textendash F bonding\cite{Robinson:2010p4582,Sato:2004p5303}, for that matter}. This provides strong evidence that the fluorine does not bond covalently to the graphene layer. The  \eli{buffer layer} peaks are still present after flourination, indicating that only partial flourination of the sample has occured.Thus the \eli{buffer} layer is transformed into a single layer of graphene by intercalation, similar to the effect of hydrogen observed in quasifreestanding samples\cite{Riedl:2009p4922, Speck:2010p5373}. Similar results (not shown) on thicker initial films have shown that the fluorine interacts with monolayers and bilayers of graphene on  \eli{buffer layer}/SiC, to produce highly $p$-doped bi-layer and tri-layer graphene respectively, by converting  \eli{their buried buffer layers} to graphene.\eli{Although a detailed interpretation of the binding energy shifts reported above is difficult without considering local field effects, a qualitative picture emerges of an interfacial fluorine layer, chemically bound to Si atoms, which is fully ionic owing to charge transfer from both the Si and from the graphene C atoms.} Comparison of the peak areas of the F 1s and the C1s graphene peak (peak G in Fig. \ref{XPS} b), and applying the correct photoemission cross-section ratio\cite{YEH:1985p5399}, gives (Graphene C)/F atomic ratio of $\sim$4.6. As we have shown that the F bonds to the surface Si not the graphene C then the important ratio is Si$_{\rm surface}$/F atomic ratio ($\sim$ 1.7), implying a 60\% flourination of the sample.

In summary, using core and valence band photoemission we demonstrate that fluorine interaction with the carbon Òbuffer layerÓ on SiC leads to the formation of a strongly $p$-doped graphene layer that is decoupled from the substrate through intercalation, shifting the Dirac crossing point to 0.79 eV above the Fermi energy. This is the strongest $p$-type doping ever observed in graphene, and the fact that the intercalated structure is stable against ambient conditions suggests its use in applications. Moreover, the doping process may be patterned, which could lead to laterally structured $n-p-n$ devices.

\begin{acknowledgments}
The Advanced Light Source is supported by the Director, Office of Science, Office of Basic Energy Sciences, of the U.S. Department of Energy under Contract No. DE-AC02-05CH11231.Work in Erlangen was supported by the ESF and the DFG through the EUROCORES program EUROGRAPHENE.
\end{acknowledgments}

%

\end{document}